\documentclass[manuscript,nonacm]{acmart}
\setcopyright{none}
\usepackage{xcolor}

\usepackage{enumitem}
\usepackage{multirow}
\usepackage{tabularx}
\usepackage{makecell}

\usepackage{caption}
\usepackage{subcaption}
\usepackage{amsmath}

\AtBeginDocument{%
  }

\begin{document}

\title[ImageTalk]{ImageTalk: Designing a Multimodal AAC Text Generation System Driven by Image Recognition and Natural Language Generation}

\author{Boyin Yang}
\affiliation{%
  \institution{Department of Engineering, University of Cambridge}
  \city{Cambridge}
  \country{United Kingdom}
}
\email{by266@cam.ac.uk}

\author{Puming Jiang}
\affiliation{%
  \institution{Department of Computing, Imperial College London}
  \city{London}
  \country{United Kingdom}}

\author{Per Ola Kristensson}
\affiliation{%
  \institution{Department of Engineering, University of Cambridge}
  \city{Cambridge}
  \country{United Kingdom}
}
\email{pok21@cam.ac.uk}

\renewcommand{\shortauthors}{Yang et al.}

\begin{abstract}
  People living with Motor Neuron Disease (plwMND) frequently encounter speech and motor impairments that necessitate a reliance on augmentative and alternative communication (AAC) systems. This paper tackles the main challenge that traditional symbol-based AAC systems offer a limited vocabulary, while text entry solutions tend to exhibit low communication rates. To help plwMND articulate their needs about the system efficiently and effectively, we iteratively design and develop a novel multimodal text generation system called \emph{ImageTalk} through a tailored proxy-user-based and an end-user-based design phase. The system demonstrates pronounced keystroke savings of 95.6\%, coupled with consistent performance and high user satisfaction. We distill three design guidelines for AI-assisted text generation systems design and outline four user requirement levels tailored for AAC purposes, guiding future research in this field. 
\end{abstract}

\begin{CCSXML}
<ccs2012>
   <concept>
       <concept_id>10003120.10011738.10011774</concept_id>
       <concept_desc>Human-centered computing~Accessibility design and evaluation methods</concept_desc>
       <concept_significance>500</concept_significance>
       </concept>
   <concept>
       <concept_id>10003120.10011738.10011775</concept_id>
       <concept_desc>Human-centered computing~Accessibility technologies</concept_desc>
       <concept_significance>500</concept_significance>
       </concept>
   <concept>
       <concept_id>10010147.10010178.10010179.10010182</concept_id>
       <concept_desc>Computing methodologies~Natural language generation</concept_desc>
       <concept_significance>500</concept_significance>
       </concept>
 </ccs2012>
\end{CCSXML}

\ccsdesc[500]{Human-centered computing~Accessibility design and evaluation methods}
\ccsdesc[500]{Human-centered computing~Accessibility technologies}
\ccsdesc[500]{Computing methodologies~Natural language generation}

\keywords{multimodal text generation, AAC, human-AI interaction}


\maketitle

\section{Introduction}
\label{section:introduction}
People living with Motor Neuron Disease (plwMND) who develop speech difficulties and motor disabilities frequently rely on augmentative and alternative communication (AAC) systems to communicate. 
They are often perceived by communication partners as passive communicators, which is closely related to individual skills~\cite{light1988interaction}. 
An AAC device that can potentially help people easily share their thoughts and life experiences may induce more active communication, helping to counteract tendencies toward learned helplessness due to speech difficulties~\cite{beukelman2020augmentative, waller2019telling,yang2023tinkerable}.
Even though many AAC devices enable users to express their wants and needs, the extremely low text entry rates typically exhibited by AAC devices impede continuous conversations with people without speech and mobility difficulties due to the breakdown of a natural flow ~\cite{todman2000rate, kristensson2020design}. 
Therefore, to keep the natural conversation flow for plwMND when communicating with people without speech and mobility difficulties, one of the design challenges is to utilize minimal operation in a short time but generate stories that contain enough information~\cite{shen2022kwickchat,cai2022context,cai2023speakfaster}. 

Storytelling is one of the most important communication activities that are used for sharing experiences, conveying information, entertaining, and connecting with others. 
It is essential to clarify that the term `story' denotes a sequence of interconnected real-world events that exhibit uniqueness or exceptional interest~\cite{waller2006communication}. 
Within the context of this paper, storytelling specifically indicates one person sharing their own experience as a monologue, that is, without interaction with the conversation partner.

These narratives may be augmented with contextual information regarding individuals' emotions and can be presented orally, differing from traditional literary narratives characterized by intricate character development and complex plot structures.

To rapidly generate stories for AAC users and allow for good controllability, two powerful design materials are potentially useful: image recognition models and large language models (LLMs). 
This is because images tend to contain richer information than plain text and can be powerful tools for storytelling because they can convey complex ideas and emotions quickly and effectively, and recent studies show a great capability of LLMs to generate text based on reduced input ~\cite{shen2022kwickchat, cai2022context, valencia2023theless}. 
By extracting information from user-selected images and integrating it with user text input, LLM-powered systems can generate stories automatically. 

However, relatively few researchers with AI system development skills conduct user studies with plwMND actively contributing to AAC technology development. As an illustrative indication, a search on Google Scholar for papers published since 2023 using the query \textit{(``augmentative and alternative communication'' OR ``AAC'') AND (``large language model'' OR ``LLM'')} yielded 875 results, in stark contrast to 17,500 results for \textit{(``education'') AND ( ``large language model'' OR ``LLM'')} and 27,000 results for \textit{``large language model'' OR ``LLM''} (accessed in April 2025), highlighting the gap in AAC-focused research leveraging LLMs.

In this study, we iteratively design, develop, and evaluate \emph{ImageTalk}, a multimodal text generation system driven by image recognition models and LLMs that can generate stories for users and analyze and compare the generated stories with the output of state-of-the-art text entry AAC systems (e.g., ~\cite{yang2023ademo}). 
ImageTalk has the potential to not only increase the keystroke savings of pure text entry AAC systems, but also improve the \emph{quality} of the generated text to allow plwMND to easily share personal stories. 

This paper provides two main contributions:
First, we present ImageTalk, a multimodal language generation AAC system developed by engaging ten proxy-users and five end users. ImageTalk achieves significant keystroke savings of 95.6\%, maintaining high user satisfaction. 
Second, we distill three design guidelines and identify four levels of cooperation goals between users and text generation systems for narrative production.

The rest of this paper is structured as follows. First, we review prior work in storytelling, image recognition, and generative AI in the domain of AAC.  Thereafter, Sections~\ref{section:system_description} --~\ref{section:results_end_user} illustrate the iterative design of ImageTalk and present findings, demonstrating the proposed design process. 
Finally, we discuss the design guidelines distilled from this work and conclude.

\section{Related Work}
\subsection{Storytelling for AAC}
The investigation of storytelling in AAC aims to broaden the communicative capabilities of non-speaking individuals with motor disabilities. 
This expansion seeks to move beyond the confines of basic requests and responses to encompass a more extensive spectrum of information exchange. 
This is especially important for children with AAC requirements to develop skills to initiate new topics and engage in storytelling. 
However, without easy access and efficient communication tools, it is challenging for them to extend interactional communication, which can lead to learned helplessness and develop a passive communicative style in the long term~\cite{basil1992social}. 

The natural language generation (NLG) system \textit{How was School Today}~\cite{reiter2009using, black2010using} shows a great achievement in helping cognitively capable children with AAC requirements to have real conversations about their day with adults, almost for their first time in real life. 
This system relies on data obtained from RFID sensors that monitor users' daily activities. These data can be acquired through two methods: either automatically, via RFID readers affixed to users' wheelchairs when they are within school premises where RFID tags are deployed, or manually, with caregivers swiping RFID cards on behalf of the users.
However, the data the system can collect is confined to the specified range of RFID tag placement.
Additionally, in many instances, caregiver assistance becomes necessary.
That is, the system cannot gather data beyond the predetermined environment.

With the evolution of mobile devices, subsequent research on storytelling for AAC has incorporated multimedia elements (e.g., photographs, audio, GPS data, etc.) to enhance the storytelling experience. 
These multimedia components serve as supplementary information, facilitating storytelling by connecting these data and presenting them to conversational partners~\cite{black2011amobile, mahmud2010xtag, rodil2018sharing}.

However, a recurring limitation within these studies pertains to the constraints imposed by the predefined narrative structure and/or the predefined environment, potentially restricting the conversational content and impeding the flow of communication.

\subsection{Image Recognition for AAC}
Image recognition technology serves to transform visual signals into text, offering particular advantages to AAC users.
This is primarily because the extracted textual representations tend to capture context information, thereby aiding AAC users in language construction. 

The ACE-LP project introduces a novel approach by incorporating multimodal sensor data, including photographs, to enhance language prediction~\cite{black2016ace}. 
Kane et al. further advance this concept by extracting context information through automatic image captioning, which recognizes objects in real-time photographs taken by the camera integrated into the AAC device. The identified objects are then listed on the text entry keyboard~\cite{kane2017lets}. 
Consequently, the system updates suggested words for users whenever the image changes. 
In more recent studies, Vargas et al. propose an innovative method that generates a rank of keywords and phrases aligned with symbols from a single user-selected photograph for symbol-based AAC systems~\cite{vargas2021automated, vargas2022aac}. 

These studies aim to enhance text entry rates of AAC users by generating vocabulary (e.g., words and phrases) from photographs.  
Nonetheless, similar to other AAC systems that provide word or phrase predictions to facilitate keystroke savings and enhance text entry rates, a fundamental design challenge persists. 
Specifically, the utility of such word or phrase prediction has limited ability to enhance keystroke savings and text entry rates. 
The extent of improvement varies depending on the capabilities of AAC users. On average, a typical AAC user achieves a text entry rate of 8 to 10 words per minute (WPM) without assistance, which can be elevated to 12 to 18 WPM with the aid of word/phrase predictive AAC systems. Nevertheless, it is noteworthy that even with this enhancement, these rates remain substantially lower than typical speaking rates, which range from 125 to 185 WPM~\cite{kristensson2020design}.
Accordingly, a more advanced design for further improvement of keystroke savings and text entry rates is desired. 

% \begin{figure*}[t]
%     \centering
%     \includegraphics[width = \textwidth]{Figures/triple_diamond.pdf}
%     \caption{The triple-diamond design process consists of three design phases. In the first phase, designers use existing design material to formulate a high-level solution-neutral problem statement and function models~\cite{kristensson2020design}. This problem statement should offer a broad design space, and the function model coupled with simulated end-users allows for rapid testing of initial ideas by exploring system parameters~\cite{kristensson2021design,yang2023imperfect}. At the end of the first phase, designers produce a set of potential designs for proxy-users in the subsequent design phase. The second design phase involves the testing of the potential design solutions with proxy-users, where implicit user performance that cannot be captured from simulations is investigated in greater depth, and a high-fidelity prototype is developed for further studies with end-users. Finally, the third phase involves end-users reliant on the actual use scenario, focusing on improving the interaction design, which closely mirrors a conventional user-centered design process, with the aim to refine and finalize the system design. }
%     \label{fig:triple_diamond}
% \end{figure*}

\subsection{Large Language Model for AAC Text Entry}
Large Language Models have great potential to address the challenge of dramatically improving keystroke savings and text entry rates by their nature. 
Recent studies, for instance, introduce a keyword-based text entry approach tailored for AAC purposes~\cite{shen2022kwickchat,yang2023ademo}.
In this method, users input keywords corresponding to their intended sentences, and the AI-powered text entry system automatically generates a completed sentence, resulting in potential keystroke savings of up to 70\%. 
Subsequent research highlights that, with dedicated interaction design, these keystroke savings can translate into an improvement in text entry rates~\cite{yang2023imperfect}. 
Valencia et al. contribute to this domain by proposing the concept of Speech Macros, serving as boundary objects and design probes to exemplify LLMs' capabilities in sentence generation for AAC applications~\cite{valencia2023theless}. 

To further leverage the capability of LLMs for AAC, we conjecture that context information extracted from photographs, coupled with keyword inputs, has the capacity to facilitate storytelling, the multi-sentence narrative, with reasonable satisfaction. 
This innovative approach holds the promise of delivering extensive keystroke savings and offers great potential for increasing text entry rates, especially when complemented by further interaction design.

\section{System Design}
\label{section:system_description}
\begin{figure*}[ht]
	\centering
		\includegraphics[width = \textwidth]{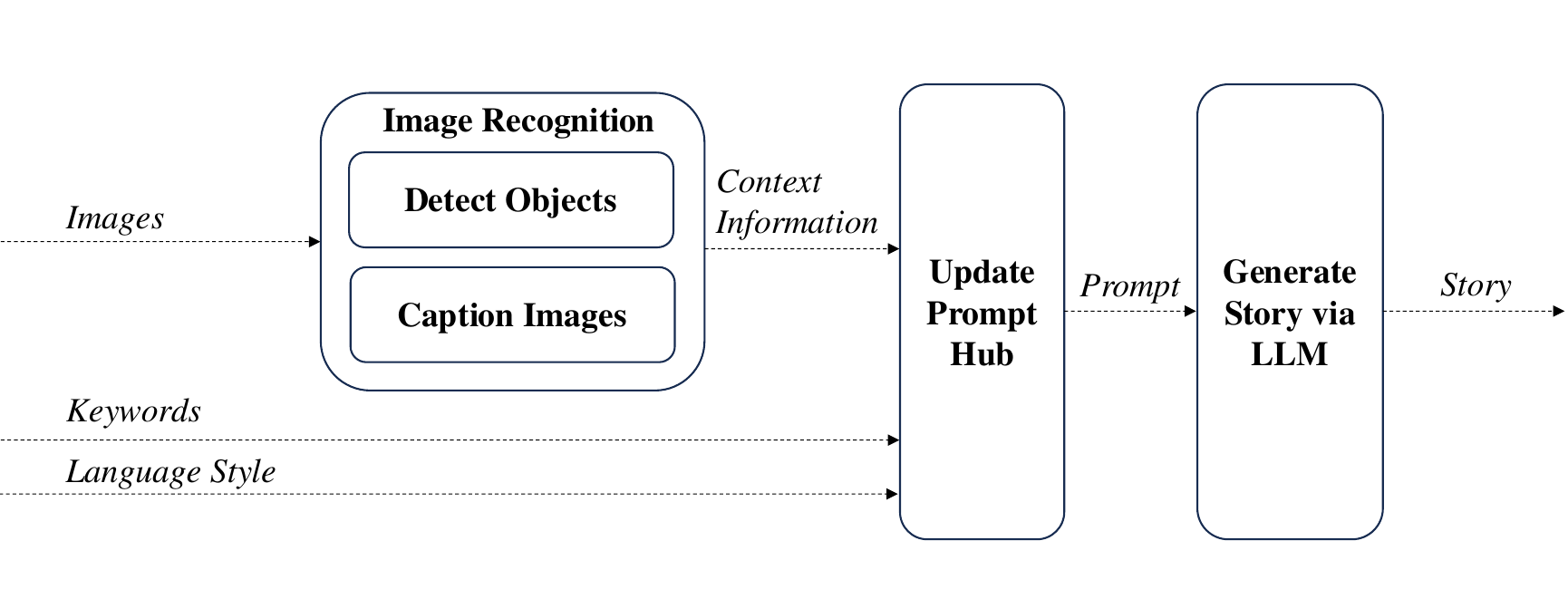} 
		\caption{The function structure model for the ImageTalk system. The fonts indicate different element types. \textbf{Bold} text along with the rectangle indicates the functions and sub-functions. \textit{Italic} text aligned with dashed lines represents the system input and output. Normal text aligned with dashed lines represents the internal information flow. Fig.~\ref{fig:workflow_example_update} shows an example workflow of this conceptual design.}  % enter the figure caption here
       \label{fig:function_structure_model} 
\end{figure*}

The function structure model allows designers to understand system functions and information flows between functions~\cite{kristensson2021design}. 
We adopt this model to illustrate the function descriptions of the multimodal text generation system, ImageTalk. 
To be more specific, this system includes three main functions: \textbf{Image Recognition}, \textbf{Update Prompt Hub}, and \textbf{Generate Story via LLM}. 
These functions are connected by signal flows represented by text with different fonts and different types of lines (see Fig. ~\ref{fig:function_structure_model}). 

The signal flows are either provided by the system or the user. 
In general, the user inputs \textit{Images}, \textit{Keywords}, and \textit{Language Style} to the system, and the system outputs a \textit{Story}. 
Throughout this process, internal data are generated, including Context Information and Prompt. 
Specifically, to maximize the context information extraction, image inputs are processed by two sub-functions of \textbf{Image Recognition}: \textbf{Detect Objects} and \textbf{Caption Images}.
This context information is transmitted to the prompt hub, along with the keywords. 
Then, the updated prompt allows the LLM to generate a story. 
This design not only allows the system to generate desired stories with a small number of user inputs, but also allows users to control the direction of generated stories by keywords. 

In reality, this system can be well integrated with existing continuous life-logs cameras such as SenseCam~\cite{hodges2006sensecam} and Automatic Alt-text~\cite{wu2017automatic} so that images can be captured automatically for AAC users.
However, capturing photographs is out of the scope of this paper. 
In this study, we focus on discussing the capability and feasibility of this novel interaction design in terms of two essential aspects for human users: keystroke savings and people's attitudes toward the generated story. 

To achieve this goal, we develop the multimodal text generation system, ImageTalk, using state-of-the-art image recognition models and LLMs. 

\subsection{Image Recognition Model in ImageTalk}
% Transformer is a state-of-the-art model. . - O

To extract different levels of context information from images, we employ both an image captioning model and an object detection model, where the former summarizes the overall information from an image while the latter extracts specific objects from the image. 
These two types of information may play varying degrees of importance in constructing a narrative.  

\begin{figure}
    \centering
    \includegraphics[width = \textwidth]{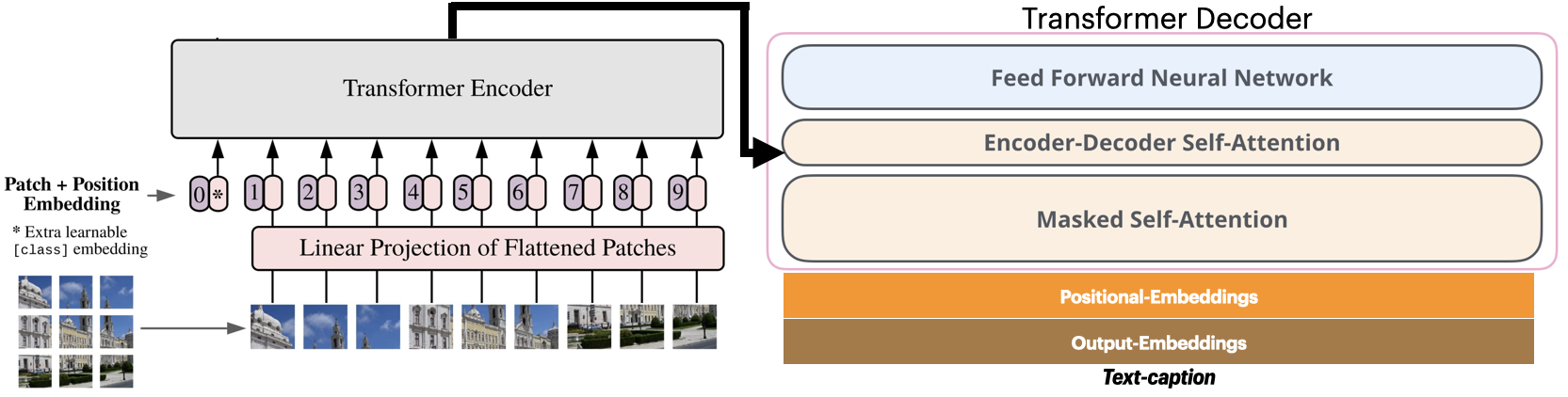}
    \caption{The architecture of the widely adopted VED model using ViT as the encoder and GPT-2 as the decoder~\cite{kumar2022imagecaptioning}.}
    \label{fig:VED}
\end{figure}

For the image captioning task, we adopt the Vision Encoder-Decoder (VED) model that employs a pre-trained image Transformer as an encoder and a pre-trained text Transformer model as the decoder~\cite{li2022trocr}. 
Figure~\ref{fig:VED} illustrates the model for our task where we adopt Vision Transformer (ViT)~\cite{dosovitskiy2021image} as the image encoder and Generative Pretrained Transformer 2 (GPT-2)~\cite{radford2019language} as the text decoder. 

\begin{figure*}
    \centering
    \includegraphics[width = \textwidth]{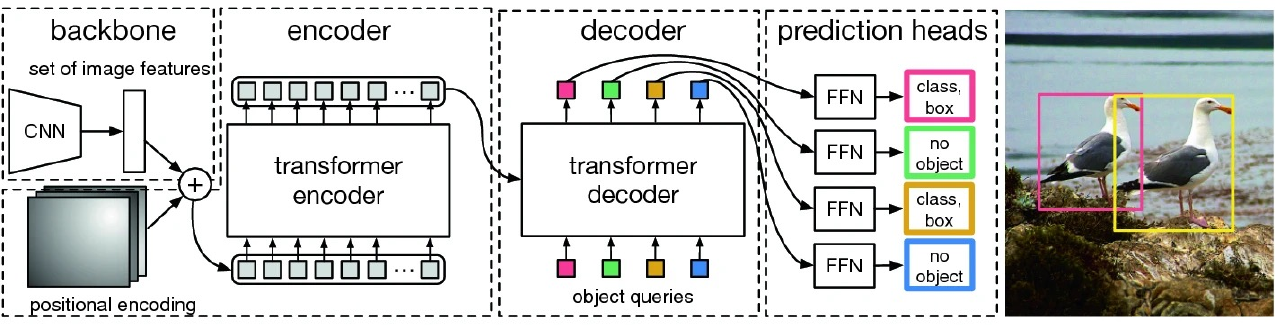}
    \caption{The architecture of DETR. This model uses a conventional neural network (CNN) backbone to learn a 2D representation of an input image. The model flattens the input image and supplements it with a positional encoding before passing it into a transformer encoder. A transformer decoder then takes as input a small fixed number of learned positional embeddings, which are called ``object queries'', and additionally attends to the encoder output. The model passes each output embedding of the decoder to a shared feed-forward network (FFN) that predicts either a detection (class and bounding box) or a ``no object'' class.~\cite{carion2020end}}
    \label{fig:detection_transformer}
\end{figure*}

For object detection, we adopt the DETR (DEtection TRansformer) model~\cite{carion2020end}. 
This is a Transformer-based end-to-end deep learning model driven by a set of learnable ``object queries'' that attend to different positions in the image, enabling the model to focus on different regions of interest for object detection in parallel with competitive accuracy and efficiency. 
Figure~\ref{fig:detection_transformer} illustrates the architecture of this model. 

The combination of these two models provides a holistic description of the visual scene in the form of a caption while also highlighting significant objects present in the images. 
The technical details are out of the scope of this paper, while these models are integrated into the ImageTalk system. The source code will be provided (in camera ready). 

\subsection{Employed LLM and Prompt in ImageTalk}
The results from our image recognition models are compiled into a corpus. This corpus, combined with user-provided keywords, is incorporated into pre-structured prompts.
The GPT-3.5 model, developed by OpenAI, is a cutting-edge language model renowned for its swift response and exceptional adaptability. It is distinguished not only by its rapidity but also by its capacity to interpret and produce human-like text across a broad range of subjects. By synergizing pre-structured prompts, image-derived information, and user-supplied keywords, we harness the capabilities of GPT-3.5 to craft stories on demand. These prompts direct the GPT-3.5 model to produce a first-person narrative, closely aligning with the given corpus and keywords while ensuring no unrelated details or information infiltrate the narrative.

\subsection{Example Workflow}

\begin{figure}[H]
\begin{center}
    \includegraphics[width = \textwidth]{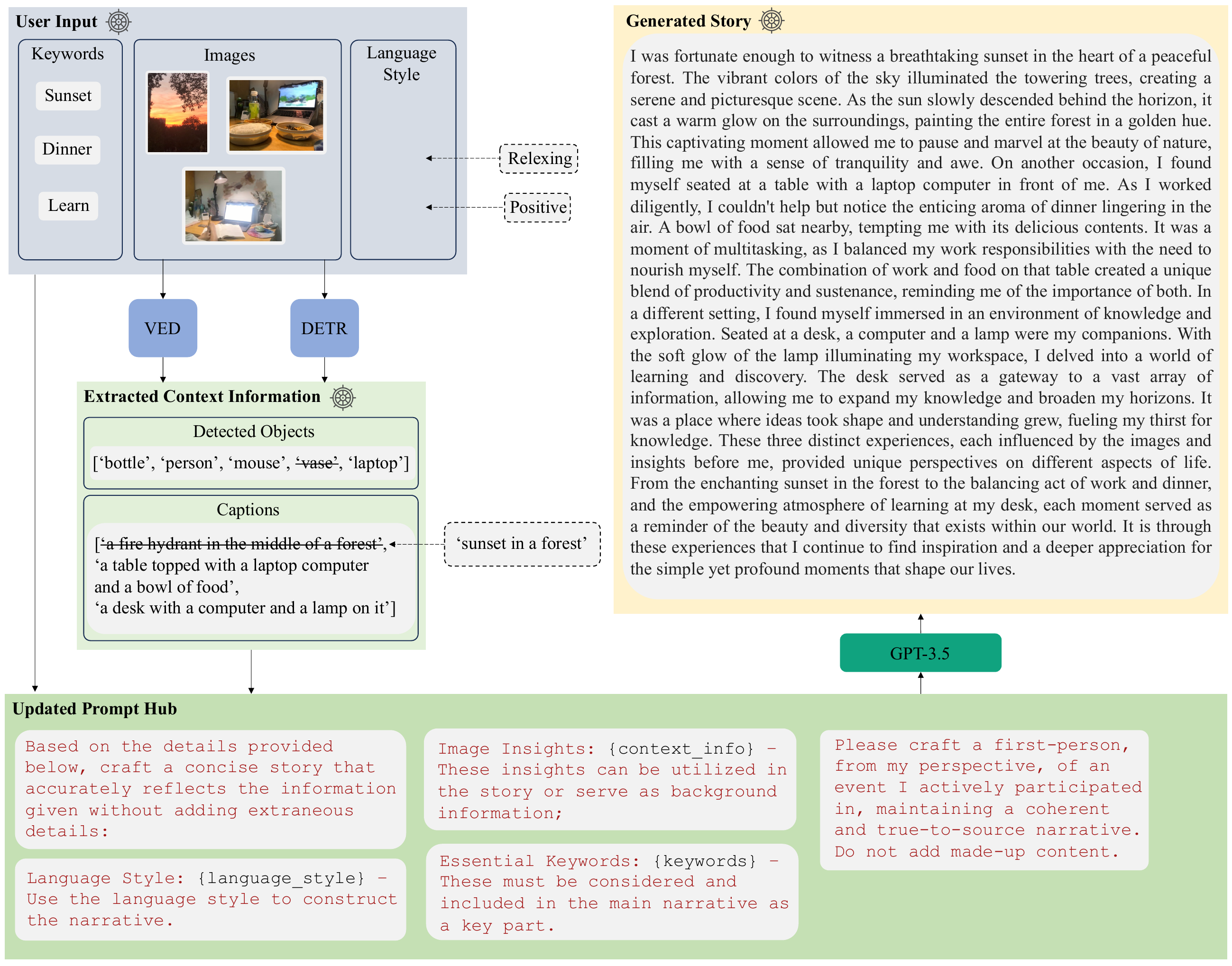} 
    \caption{An example workflow of the ImageTalk system. The solid line delineates the sequential steps within the workflow, commencing with user input. At this initial stage, users select images and input keywords, along with language styles corresponding to the intended narrative. These selected images undergo simultaneous processing through an image captioning model, VED~\cite{li2022trocr}, and an object detection model, DETR~\cite{carion2020end}. Extracted context information from images is subsequently transmitted to the prompt hub, together with the user-provided keywords. The updated prompt is relayed to the LLM, GPT-3.5, as employed in this system. The LLM generates a narrative that encapsulates the selected images and user-inputted keywords. The \textit{steering wheel} symbol demotes steps that allow for direct user manipulation, with the system responding to these operations in real-time. Specifically, the strike-through and the dashed boxes and arrows indicate the user's steering.} 
    \label{fig:workflow_example_update} 
\end{center}
\end{figure}

\begin{figure}[H]
    \centering
    \includegraphics[width = \textwidth]{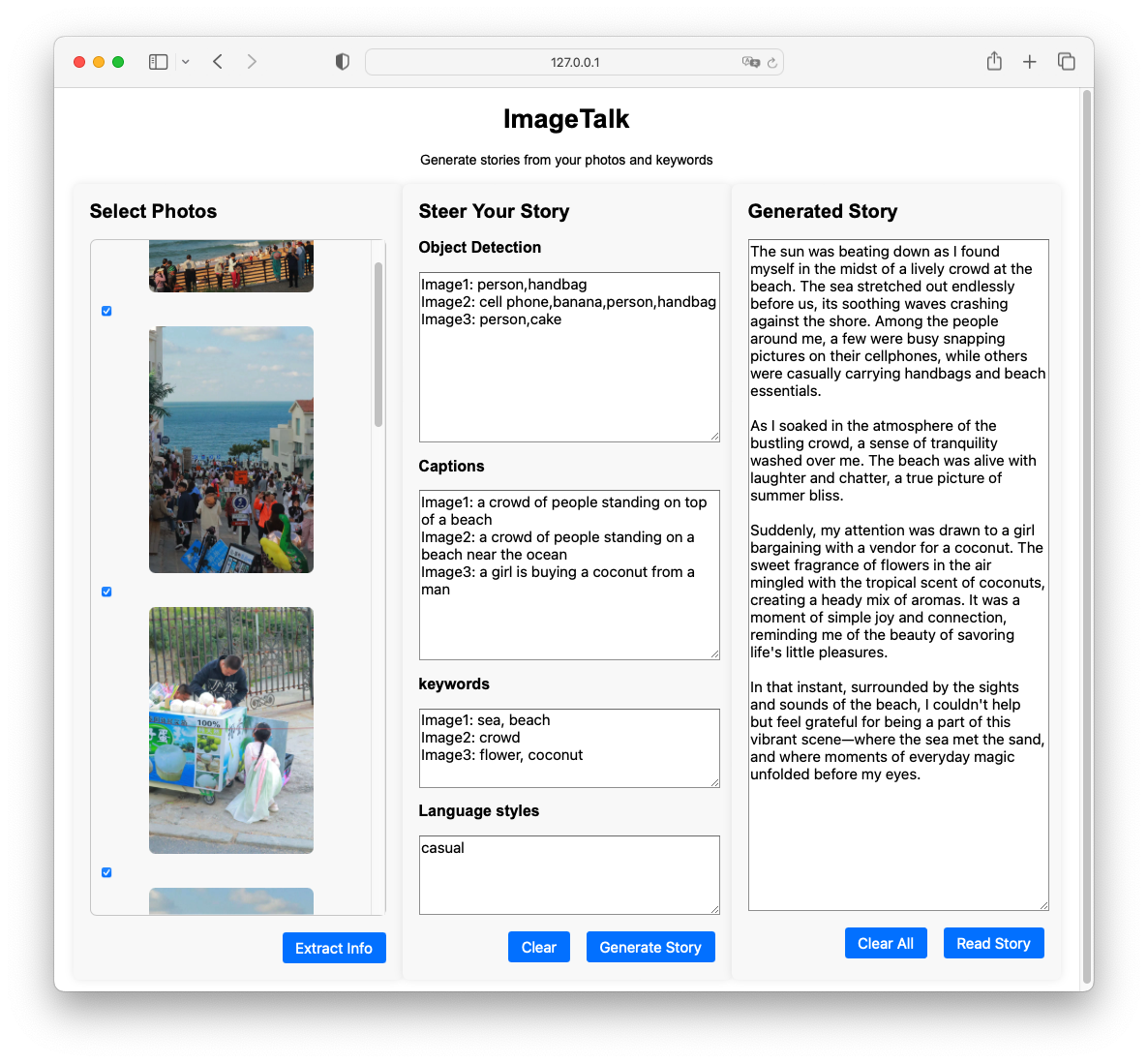}
    \caption{This ImageTalk user interface is used for functional capability evaluation, operated by the researcher on behalf of the participants. Before the user study begins, participants provide images to the researcher, who then uploads them into the system. During the study, the researcher selects images supplied by the participant. The system extracts context information from these images and displays it in the Object Detection and Captions areas. The participant can instruct the researcher on any desired edits to this generated information. Subsequently, the participant provides keywords associated with each selected image and specifies the desired language style. After the researcher inputs this information and clicks the Generate Story button, the system automatically creates a story. It is important to note that this version of ImageTalk is NOT the finalized system but a transitional design and experiment tool aimed at evaluating the concept and participants' acceptance of the generated context. This GUI helps us better understand the design space, and the finalized system will require further iterations of the interaction design to meet end-users' operational needs, which will be addressed in future work. }
    \label{fig:image_talk_gui}
\end{figure}

Figure \ref{fig:workflow_example_update} outlines the ImageTalk system's workflow created in the third designed phase, commencing with user-selected images, keywords, and language styles. These images are processed using the VED~\cite{li2022trocr} and DETR~\cite{carion2020end} models. Contextual information extracted combines with user keywords and expected language styles in the prompt hub and is then passed to the LLM, GPT-3.5, for narrative generation. 
To ensure the stories are generated as the user expected, the system allows for a combination of automation and human manipulation. 
Figure~\ref{fig:image_talk_gui} shows a GUI for function capability evaluation. It is worth mentioning that this graphic user interface (GUI) is a functional design throughout the whole triple diamond design process while not necessarily reflecting the finalized system GUI, which may also consider usability and how users interact with the interface.

This workflow is delineated using participant EndUserP1's example as a representative demonstration of the overall process from our evaluation, which we will present later in this paper. Private information on the images is masked. For more diverse content examples, please refer to Section~\ref{section:Examples} and Section~\ref{section:Examples_stage2}.

\section{Evaluation}
Given the nature of the limited communication rate and physical capabilities of AAC users, it is incredibly challenging for AAC users to participate directly in the evaluation of a system that may not have a dedicated AAC user interface. 
As indicated in prior work (e.g.~\cite{kristensson2020design}), it is extremely difficult to validate a system with AAC users at an early stage of the design process, and this challenge interferes with established user-centered design practices. 
Nonetheless, there are still many aspects that can be evaluated without the involvement of AAC users at this early stage of the design process. 
For example, text generation strategies and context-aware sentence retrieval features can be quantitatively evaluated via envelope analyses at the early design stage (e.g.~\cite{kristensson2020design, kristensson2021design, yang2023imperfect}). 
This evaluation strategy reduces the time spent in the early human-centered design and system development cycle by offering a refined and more informed suggestion of a system design prior to any actual studies with AAC users. 

Considering the nature of storytelling, the fundamental desire for users without speech and mobility difficulties and AAC users remains the same: sharing expected information and/or feelings with conversation partners. 
Therefore, to gain fundamental design inspirations of ImageTalk for further user study with AAC users, we carried out studies with ten users without speech and mobility difficulties during the human proxy-based design phase that aims to evaluate the preliminary keystroke savings and people's attitudes toward AI-generated stories based on photographs and keywords.
Building upon the findings from this phase, we refined the system design by adding a steering function. 
During the end-user-centered design phase, we engaged five participants living with MND to validate the triple diamond design process and assess whether this new design iteration lead to an improved system for AAC end-users.

\subsection{Method}
We evaluated the quality of I-generated stories by ImageTalk by comparing them with reference stories provided by the human participants, the stories generated via the pre-existing Keywords-to-Story (KTS) approach on an existing AI-powered text entry system, Tinkerable-AAC~\cite{yang2023tinkerable,yang2023ademo}. 

\subsubsection{Participants}
In the proxy-user-based design phase, we recruited ten participants using convenience sampling. 
The participants were all individuals without speech and mobility difficulties aged between 20 and 35. 
Based on self-reporting, seven participants described themselves as fluent English speakers, two as intermediate English users, and one as a native English speaker.
In terms of prior experience with GenAI, five participants reported having used GenAI at least once before this study, three reported using GenAI every week, and two reported integrating GenAI into their daily workflow. 

In the end-user-centered design phase, we recruited five participants using convenience sampling. 
These participants all lived with motor neurone disease (MND), and their ages ranged between 20 and 45.
Based on self-reporting, they all face challenges in daily text entry tasks. 
One participant had experience with AAC devices, while four had no such experience.
In terms of language ability, one participant described themselves as fluent English users, three as intermediate English users, and one as basic English users. 
In terms of prior experience with GenAI, four participants reported using GenAI every week, and one reported having used GenAI at least once before this study.

\subsubsection{Procedure: Proxy-User-Based Design Phase}
\label{section:procedure}
In the proxy-user-based design phase, the evaluation aims to answer the question  ``Does ImageTalk have the potential to increase AAC users' communication rate?'', which can be further decomposed into two main questions that guide our design: 

\begin{itemize}
    \item To what extent can the multimodal text generation system ImageTalk increase keystroke savings? 
    \item How do human users perceive the sentences generated by AI for storytelling purposes? 
\end{itemize}

To address these questions, the study was structured in two parts. 
In the first part the proxy-users were instructed to provide a set of photographs and a set of keywords pertaining to a brief narrative recounting recent events in their lives. 
Each proxy-user was subsequently tasked with manually typing out this narrative. 
The researcher (i.e., the author) then used these provided elements to generate two distinct narratives for each proxy-user. 
This generation process employed two distinct approaches: the Keywords-to-Story (KTS) approach, as detailed in Section~\ref{section:benchmark}, serving as the benchmark, and the multimodal text generation system ImageTalk, as introduced in Section~\ref{section:system_description}.
The KTS approach relied solely on the provided keywords, while ImageTalk used both the provided photographs and keywords. 
We used this design to ensure that proxy-users evaluated the generated narratives without potential bias resulting from the interaction experience with both systems.
The second part of the user study involved a semi-structured interview. During this session, proxy-users encountered two AI-generated narratives and were invited to share their impressions. Specifically, they were asked to identify the accuracy and language style of AI-generated stories and to elaborate on their reasons for these preferences. The proxy-users remained uninformed about the specific methodologies employed in generating the narratives.

\subsubsection{Procedure: End-User-Centered Design Phase}
In the end-user-centered design phase, the evaluation aims to answer the question ``How do AAC users interact with the system to generate expected stories?'', which can be further decomposed into three main questions:
\begin{itemize}
    \item Does the system output perform consistently between proxy-users and end-users?
    \item What are the interaction points for users to steer the generated story?
    \item How effective is the design of allowing users to steer the generated story?
\end{itemize}

To address these questions, we carried out a three-part study with five plwMND who have speech and mobility difficulties. 
The first two parts of the study were identical to the parts used in the proxy-user-based design phase, and we used these parts of the study to verify that the findings in the proxy-user design stage were consistent with the findings in the end-user design stage.
At the end of the interview in part two, we asked the end-users to provide editing suggestions for the ImageTalk story in the form of a set of keywords. In part three of the study, we generated a new story for each end-user. We carried out a brief interview immediately afterward, asking about the level of satisfaction experienced by the end-user in response to the new story compared to the previously generated version. 

\subsubsection{System Settings and the Benchmark: Keywords-to-Story}
\label{section:benchmark}
Prior research has demonstrated that keyword-based text entry methods implemented within AI-powered text entry systems hold the potential to significantly enhance keystroke savings while maintaining the quality of generated sentences (e.g.~\cite{shen2022kwickchat}). 
Subsequent work further improves such interaction by introducing an open-source AI-powered text entry system, Tinkerable-AAC, designed for customization and experimentation~\cite{yang2023ademo}. 

For our study, we adopted this latter system for the KTS task, serving as our benchmark. 
To ensure methodological consistency and control over variables, we selected the same LLM engine, GPT-3.5, for both ImageTalk and KTS.
Additionally, we standardized the parameter values of the LLM across both narrative generation tasks. 
Moreover, we turned off the word prediction functions on Tinkerable-AAC for the same purpose.

\subsection{Metrics}

\subsubsection{Keystroke Savings}
Keystroke savings is an established metric that characterizes the extent to which a system's design reduces the number of keystrokes or user interactions necessary to complete a task. 
We calculate keystroke savings for generated narratives through both multimodal input and exclusive keyword input. 

We determine the percentage of keystrokes saved compared to the total number of characters the user would be required to manually type without the aid of generative functions. 
It is essential to clarify that word prediction functions, commonly integrated into many AAC text entry systems, were deliberately deactivated for this study. 
This deactivation implies that the researcher had to manually type each keyword letter by letter.
To calculate keystroke savings for both the KTS and ImageTalk methods, we use the following equation:
\begin{equation}
    KS = \frac{N_{story}-N_{keyword}}{N_{story}}\times 100\%,
\end{equation}
where $N_{story}$ represents the character count of the generated story, with punctuation omitted for simplicity, and spaces are considered as individual keystrokes; and $N_{keyword}$ is the total number of characters manually typed by the user, including spaces inserted after each keyword. 

This calculation also aims to explore the maximum achievable keystroke savings through the use of ImageTalk. 
Thus, our calculation of keystroke savings for ImageTalk deliberately excludes the additional step of photograph selection. This methodological choice enables a focused assessment of the inherent keystroke-saving capabilities of ImageTalk, independent of the specific photograph selection process.

\subsubsection{Semantic Similarity} 
\label{section:semantic_similarity}

Semantic similarity refers to a measure of how similar two pieces of text are in terms of their meaning rather than just their surface form or syntactic structure. 
Word2Vec~\cite{mikolov2013efficient}, a commonly used word embedding technique, encapsulates semantic relationships among words, effectively capturing nuances like synonyms and semantic analogies. This is achieved by representing words in representing them as vectors in a high-dimensional vector space, wherein the semantic proximity between words is mirrored by the geometric proximity of their respective vectors. For our purposes, we use a Word2Vec model trained on the expansive Google News dataset ~\cite{google2013word2vec}, which maps each word to a 300-dimensional vector based on its meaning.

To obtain the semantic vector representation of an entire document or story, we average the embeddings of its individual words. Mathematically, \(w_1, w_2, \ldots, w_n\) are the word vectors of a story, and the document's semantic representation \(D\) can be derived as:

\[ D = \frac{w_1 + w_2 + \ldots + w_n}{n} \]

We measure the cosine similarities between these average story vectors. This metric represents the cosine of the angle between the respective vectors. A higher cosine similarity score implies a narrower angle, denoting a higher degree of semantic resemblance between the stories in comparison. Perfectly matching narratives would yield a cosine similarity of one.

\section{Results: Proxy-User-Based Design Phase}
\subsection{Examples}
\label{section:Examples}
Table~\ref{table:output_examples} shows examples of user inputs, reference stories written by proxy-users, and corresponding system outputs generated by ImageTalk and the KTS method. 
The proxy-users were instructed to freely select single or multiple photographs and provide necessary keywords that assist in constructing the reference story they provided. 
As evident in the table, ImageTalk adeptly integrates keywords and information extracted from the photographs into the generated narratives, and KTS also incorporates keywords into its output.
In general, ImageTalk tends to produce narratives enriched with detailed information, whereas the KTS approach tends to generate more concise narratives.
These examples have been chosen due to the diverse use of keywords and photographs by users and typical errors generated by AI (see Section~\ref{section:design_guidelines}).

\begin{table}
    \centering
    \includegraphics[width = \textwidth]{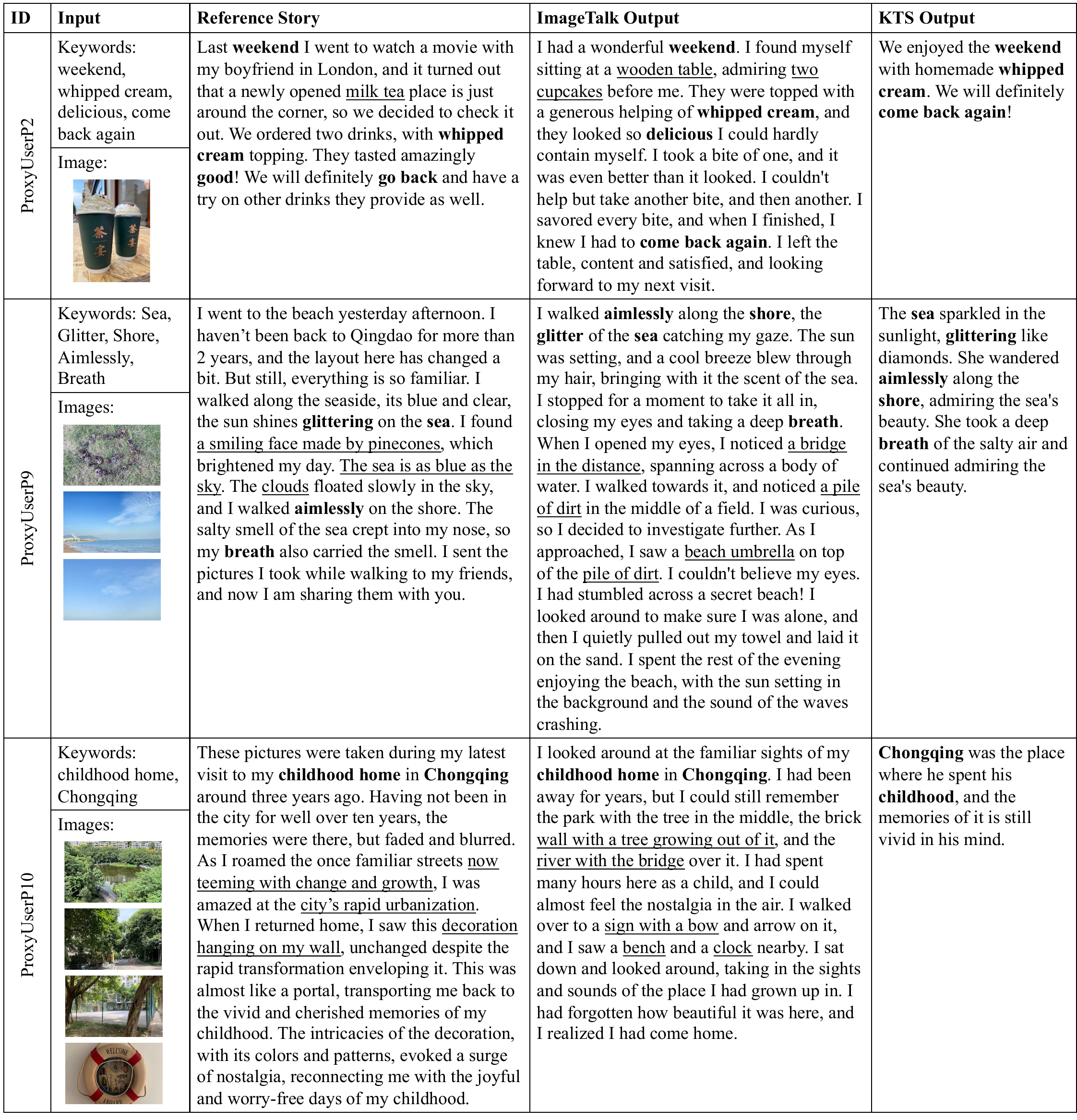}
    \caption{Three typical examples of proxy-user inputs, reference stories, and corresponding system outputs generated by ImageTalk and KTS methods. Keywords are highlighted in bold; context information extracted from the image is underscored. }
    \label{table:output_examples}
\end{table}

\subsection{Keystroke Savings}
% O

\begin{table}[h]
\begin{tabular}{lccll}
                              & \multicolumn{1}{l}{Mean} & \multicolumn{1}{l}{Standard Deviation} &  &  \\ \cline{1-3}
Keystroke savings for KTS & 66.5\%                  & 14.5\%                                 &  &  \\
Keystroke savings for ImageTalk       & \textbf{94.4\%}                  & \textbf{3.3\%}                                &  &  \\ \cline{1-3}
Keywords/Reference story      & 10.1\%                  & 4.2\%                                 &  &  \\ \cline{1-3}
\end{tabular}
\caption{Proxy-users' average keystroke savings along with the corresponding standard deviations for both ImageTalk and KTS approaches. Additionally, the third row displays the keyword count relative to the reference story.}
\label{table:keystroke_savings}
\end{table}

Table \ref{table:keystroke_savings} presents an overview of the mean and standard deviation of keystroke savings observed in both the ImageTalk and KTS methods. 
Notably, even though both methods make use of the same set of keywords, the ImageTalk method demonstrates significantly superior keystroke savings, achieving a remarkable 42\% improvement compared to the KTS approach.
Moreover, it is noteworthy that ImageTalk exhibits a higher degree of consistency in terms of keystroke savings, boasting a level of stability approximately 4.4 times greater than that of the KTS approach, as measured by the standard deviation in keystroke savings. 

In addition, the third row in Table \ref{table:keystroke_savings} delineates the ratio of keystrokes required to type the provided keywords as opposed to the complete reference story, emphasizing the effectiveness of using keywords instead of the full reference story in general.

\subsection{Quality of AI-Generated Sentences}
To obtain a comprehensive assessment of the quality of AI-generated sentences, our evaluation approach consists of both quantitative measurements of semantic similarity and qualitative human evaluations.

\subsubsection{Semantic Similarities}
\begin{figure}[h]
    \begin{center}
        \includegraphics[width = 0.6\textwidth]{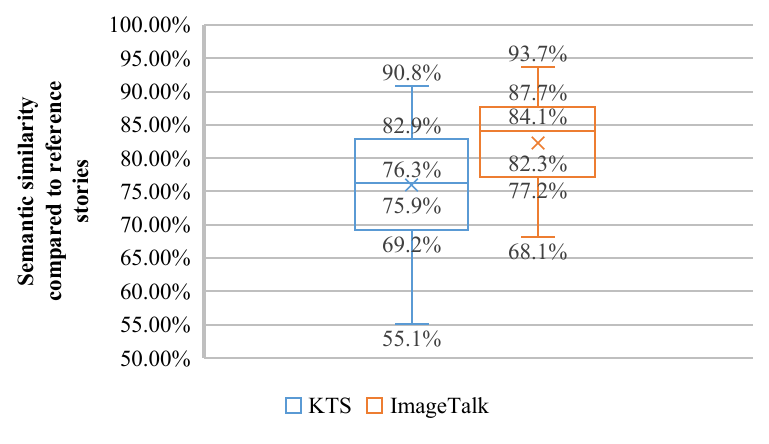} 
        \caption{Proxy-users' box plot that illustrates the semantic similarity between the generated stories and the original data. The mean is denoted by a cross. The box plots themselves capture the range from minimum to maximum values, with lines indicating the two quartiles and the median. The orange error bar corresponds to ImageTalk, which consistently exhibits higher semantic similarity compared to the blue bar, representing the benchmark KTS method.} % enter the figure caption here
       
        \label{fig:semantic_similarities} 
    \end{center}
\end{figure}

We use Word2Vec (see Section~\ref{section:semantic_similarity}) to calculate the semantic similarity between stories generated by ImageTalk and their corresponding reference stories, and we use 
the same method is adopted to calculate semantic similarity between stories generated by KTS and their corresponding reference stories, which serves as the benchmark. 
The results are shown in Figure~\ref{fig:semantic_similarities}.

It is evident that ImageTalk consistently achieves higher similarity scores for each proxy-user, with an overall average of 82.3\% compared to the KTS method's average of 75.9\% and a more consistent performance. 
This implies that the narratives generated by ImageTalk incorporate words that bear a closer semantic resemblance to the reference stories. Nevertheless, we acknowledge that this quantitative calculation may not fully capture the human perception of sentence quality. 

\subsubsection{Human Evaluation}
\begin{figure}[h]
    \begin{center}
        \includegraphics[width = 0.6\textwidth]{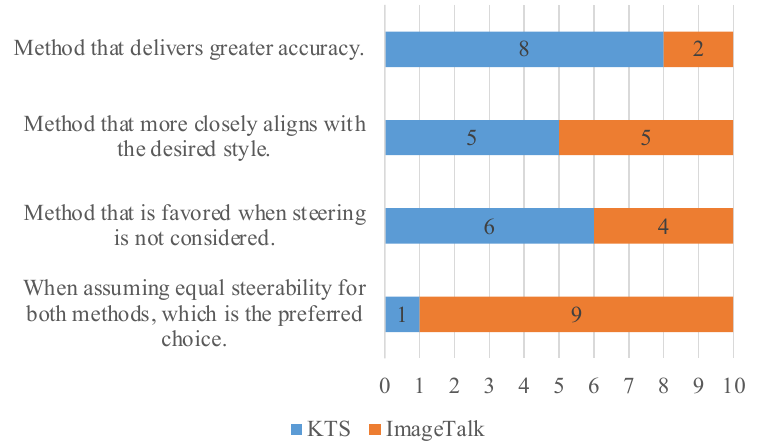} 
        \caption{Proxy-users' responses to binary-choice questions during the interview.} 
        \label{fig:binary_choice} 
    \end{center}
\end{figure}

Figure~\ref{fig:binary_choice} presents the responses of ten proxy-users to binary-choice questions posed during the interview, inquiring about their personal preferences and perceived semantic accuracy of the narratives generated by both ImageTalk and KTS methods.

Notably, the narratives produced by both methods exhibit similar preferences by overall proxy-users. 
Moreover, the language styles of both methods are equally preferred. 
However, it is interesting that, despite the semantic similarity data suggesting otherwise (see Figure~\ref{fig:semantic_similarities}), the majority (eight out of ten) of proxy-users expressed a preference for the perceived accuracy of the narratives generated by the KTS approach in comparison to ImageTalk, relative to their own reference stories.
Even more interestingly, in a scenario where both generated narratives were assumed to provide an equal degree of steerability, the preference shifted significantly. 
Nine proxy-users favored editing the narrative generated by ImageTalk. 
This mismatch opens up opportunities for design, which we distill into three design guidelines for such systems.

\subsection{Design Guidelines}
\label{section:design_guidelines}
We analyzed the rationales behind the mismatch of semantic analyses and human evaluation results, evidenced by proxy-users' feedback, which opens up at least three essential human-in-the-loop design insights for improving AI-powered multimodal text generation systems.

\subsubsection{Support Fine-Grained Language Style Control}
ImageTalk demonstrates a surprisingly good performance in keystroke savings and receives plenty of positive feedback from proxy-users about language style. 
For example, those who favored the ImageTale model described it as 
\begin{quote}
    \textit{``giving a sense of sharing life experiences''} -- ProxyUserP1, ProxyUserP4; 
\end{quote}
\begin{quote}
    \textit{``being more authentic and colloquial''} -- ProxyUserP6, ProxyUserP8, ProxyUserP9;
\end{quote}
\begin{quote}
      \textit{``presenting in the first person''} -- ProxyUserP9. 
\end{quote}

This is especially true when the story itself is captivating, such as a memorable trip (ProxyUserP1, ProxyUserP6). Proxy-users are then more receptive to an embellished narration.
However, when the shared events are more mundane, proxy-users may find an excessively adorned tone to be discomforting.
For example, in this scenario, proxy-users tend to describe ImageTalk as 
\begin{quote}
    \textit{``overly exaggerated''} -- ProxyUserP2, ProxyUserP5, ProxyUserP7, 
\end{quote}
while believe KTS offered a more 
\begin{quote}
    \textit{`lifelike feel'} -- ProxyUserP3, ProxyUserP10.
\end{quote}

To retain favorable system performance characteristics beyond the nature of ImageTalk, which has high keystroke savings, it becomes imperative for the system to possess an understanding of the desired language style in its generated output. 
Within the framework of human-in-the-loop design, this assumes a critical role as an interaction point, enabling human users to exercise delicate control over the language style. 
Leveraging the capabilities of LLMs, such control can be readily achieved through the modification of prompts in combination with a tailored user interface design that allows users to appropriate systems for their purposes.

\subsubsection{Detect Potential Decisive Errors in the Internal Information Flow of the System} 
A \emph{decisive error} is an error that has a significant negative impact on an outcome. 
In this case, the result of a decisive error is the generated story being unacceptable by the user. 
When considering the semantic accuracy of stories generated by AI-powered text entry systems, a decisive error can happen in two cases. 

First, the system generates erroneous information about the proxy users using the keyword and/or context information. 
This is one of the main reasons for the mismatched results. 
For example, participant ProxyUserP4 provides keywords ``Japanese'', ``food'', and ``dinner with friends'' and is meant to describe an activity that happened in a Japanese restaurant. 
ImageTalk crafted a story stating, ``I was in Japan to have dinner with my friends...''. 

Second, the system fails to identify unimportant information and generates irreverent content that subsequently influences the output. 
For example, ProxyUserP9 supplied an image of a bridge but had no intention of mentioning that bridge. 
However, the image captioning model generated the description ``a bridge over a body of water'', subsequently influencing the story.

Decisive errors have the potential to manifest in the early stages within the information flow of the AI system. 
If these errors are not promptly corrected, they tend to propagate throughout subsequent stages, exacerbating the overall impact.
Hence, it becomes essential to facilitate human user intervention at various stages of the AI system's narrative generation process. 
As an illustrative example, permitting users to select their preferred context information and furnishing prediction options for keywords endows human users with greater control over the direction of the generated narrative. 
This measure aims to preempt the occurrence of decisive errors.

\subsubsection{Support Fast Amendment of Decisive Errors in System Output}
Proxy-users' attitudes to the generated story can shift dramatically if the system output can be amended. 
A notable rationale for this shift in preference was provided by ProxyUserP6: 
\begin{quote}
    \textit{``(In ImageTalk's story,) there are moments where the time-related details don't match my expectations. Yet, the overarching story remains intact, and the inclusion of emotive sentences enhances the overall narrative.''}
\end{quote}

In essence, decisive errors often tend to be associated with specific segments of the output rather than affecting the entire result. Importantly, even within the context of steering the system towards the generation of desired narratives, decisive errors can still occur within the final system output.
Consequently, it becomes necessary to perceive the system output as a composite of multiple constituent parts, each of which is interactive and amenable to rapid amendment. Such amendments should ideally leave undisturbed those segments of the output that already meet user satisfaction criteria.
Furthermore, this amendment process can be augmented by AI capabilities that offer predictive options.

\section{Results: End-User-Centered Design Phase}
\label{section:results_end_user}
During the proxy-user-based design phase, we realized that although ImageTalk with full automation (i.e., automated ImageTalk) holds substantial potential to achieve high keystroke savings, it does not provide a consistent user satisfaction in response to a generated story. 
As a result, we distilled three guidelines, indicating the necessity of human intervention throughout the entire information flow. 
Subsequently, this system was refined with semi-automation to provide users with more control (i.e., Semi-automated ImageTalk), and it was tested with five end-users. 
The refined functions are depicted in Figure~\ref{fig:workflow_example_update}, denoted by the \textit{steering wheel symbol}. Semi-automated ImageTalk allows users to directly manipulate user input and extract context information by adding, removing, or modifying the text.
In addition, the final generated story also allows for further editing, empowering the user to be in complete control over the result.

\subsection{Examples}
\label{section:Examples_stage2}
Figure~\ref{table:output_examples_update_p13} shows an example of participant EndUserP3 interacting with Semi-automated ImageTalk: their inputs, the initial generated story, user steering, and the final generated story. 
In the study, the end-users were instructed to freely select single or multiple images, along with necessary keywords that assist in constructing the reference story they provided. The system automatically generates an associated story, as shown in the table. Then, the end-user is asked for feedback about the generated story and shares their thoughts on how they want this preliminary story to be edited. Based on this steering, the Semi-automated ImageTalk system re-generates a new story for the user. In general, all end-users were either more satisfied (four out of five end-users) or remained at a similar satisfaction level (one out of five end-users) after steering.

\begin{table*}[h]
    \centering
    \includegraphics[width = \textwidth]{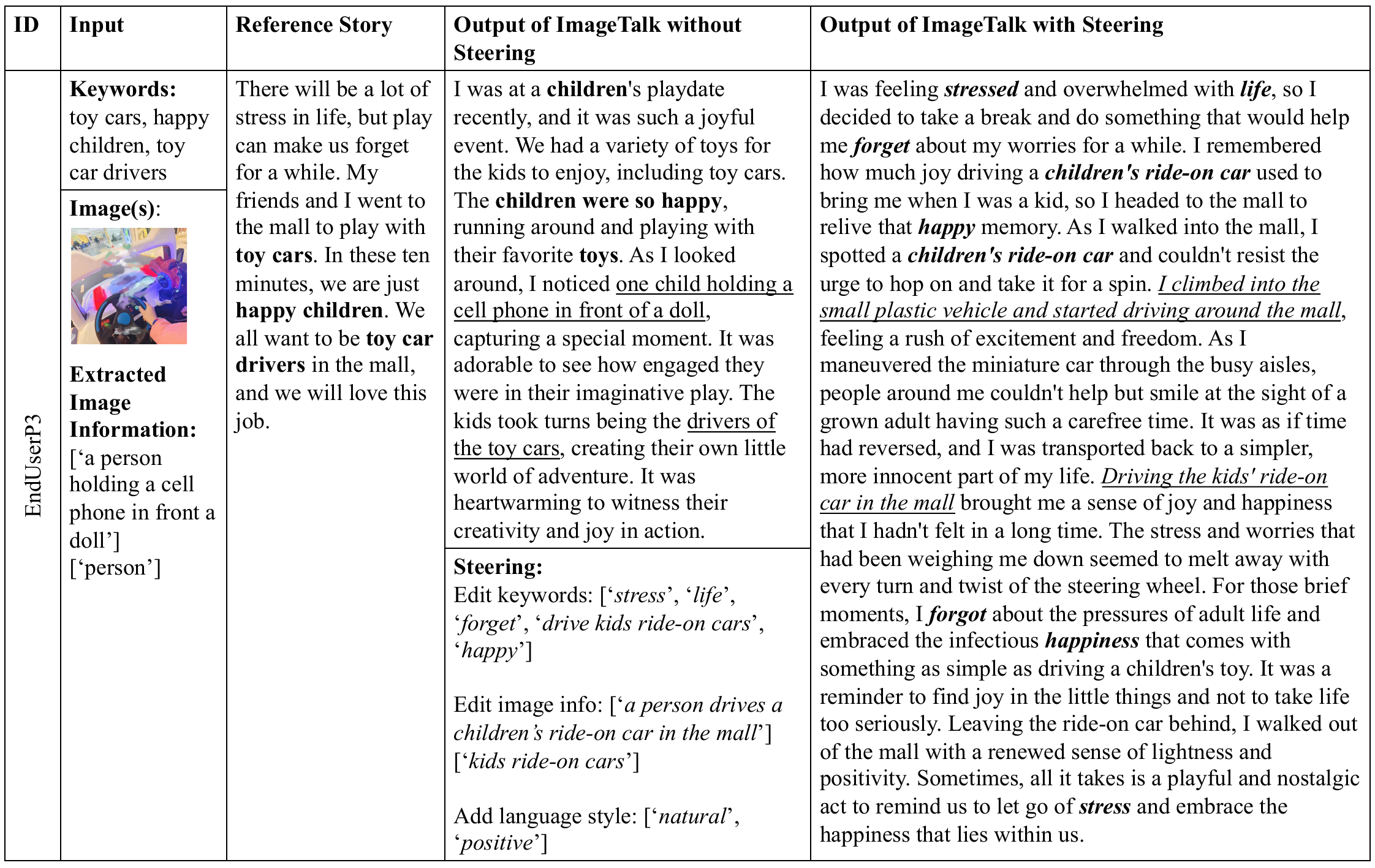}
    \caption{An example of an end-user's inputs with two outputs. The output of Automated ImageTalk (without Steering) is based on Keywords and Images. The \textbf{Extracted Image Information} is shown to the user to allow steering. The output of Semi-automated ImageTalk (with steering) is based on the updated input. }
    \label{table:output_examples_update_p13}
\end{table*}

\subsection{Keystroke Savings}
Table~\ref{table:keystroke_savings_stage2} shows an overview of the mean and standard deviation of keystroke savings in KTS, Automated ImageTalk, and Semi-automated ImageTalk methods with end-users. Notably, Automated ImageTalk and Semi-automated ImageTalk show comparable high keystroke savings and low standard deviation in keystroke savings, indicating the consistent performance of this system, which in turn yields markedly better performance than the KTS approach. 
\begin{table}[h]
\begin{tabularx}{\textwidth}{Xcc}
                                               & Mean           & \makecell{Standard \\ Deviation} \\ \hline
Keystroke savings for KTS                      & 75.6\%          & 13.6\%              \\
Keystroke savings for automated ImageTalk      & \textbf{96.4\%} & 3.1\%               \\
Keystroke savings for semi-automated ImageTalk & 95.6\%          & \textbf{3.0\%}      \\ \hline
Keywords/Reference story                       & 8.6\%           & 8.8\%               \\ \hline
\end{tabularx}
\caption{End-users' average keystroke savings along with the corresponding standard deviations for KTS, Automated ImageTalk, and Semi-automated ImageTalk approaches. Additionally, the fourth row displays the keyword count relative to the reference story.}
\label{table:keystroke_savings_stage2}
\end{table}

\subsection{Quality of AI-Generated Narratives After User Steering}
\subsubsection{Semantic Similarity}
\begin{figure}[h]
    \centering
    \includegraphics[width = 0.6\textwidth]{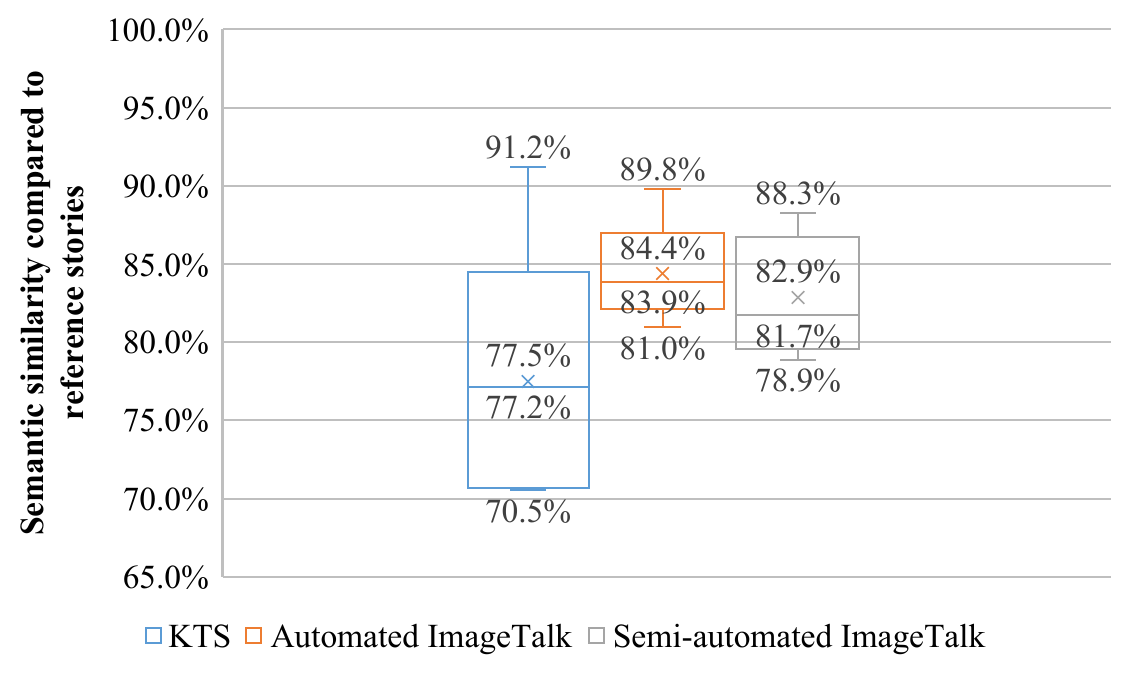}
    \caption{End-users' box plot illustrating the semantic similarity between the generated stories and their original data. The mean is denoted by a cross. The box plots themselves capture the range from minimum to maximum values, with lines indicating the two quartiles and the median. The orange error bar and gray error bar correspond to Automated ImageTalk and Semi-automated ImageTalk respectively, which exhibit higher average semantic similarity compared to the blue error bar, representing the benchmark KTS method.}
    \label{fig:semantic_similarity_stage2}
\end{figure}

To evaluate the system performance in accuracy before and after the redesign, we adopt the same approach used in Section~\ref{section:semantic_similarity} to calculate the semantic similarity between stories generated by KTS and their corresponding reference stories, Automated ImageTalk and their corresponding reference stories, and Semi-automated ImageTalk and their corresponding reference stories, respectively. 
Figure~\ref{fig:semantic_similarity_stage2} shows that both Automated ImageTalk and Semi-automated ImageTalk consistently achieve higher similarity scores for each end-user, with an overall average of 84.4\% and 82.9\%, respectively, compared to the KTS method's average of 77.5\%. 
Semi-automated ImageTalk shows a slight decrease in semantic similarity to Automated ImageTalk. We list the following observations that may explain this result: 
\begin{enumerate}
    \item The participant removes images and extracted information from images during steering (EndUserP2).
    \item The participant rephrases the keywords during steering (EndUserP3).
    \item The participant changes the intention after having read the initial generated story (EndUserP4).
    \item The participant adds keywords that are not from the reference story (EndUserP5).
\end{enumerate}

\subsubsection{User Satisfaction}
The decrease in semantic similarity after steering does not reflect user satisfaction. 
Instead, user satisfaction is greatly improved after steering. 
Figure~\ref{fig:human_evaluation_stage2} presents the responses of five end-users to binary-choice questions to the automated ImageTalk and ternary-choice questions to the semi-automated ImageTalk posed during the interview, inquiring about their personal preferences and perceived accuracy. 
As is evident in Figure~\ref{fig:human_evaluation_stage2} b), the semi-automated ImageTalk outperforms the automated ImageTalk in both accuracy and language style. 

We carried out two amendments of the system based on our observation of end-user steering: (1) an ability to correct misidentified information from images (EndUserP1, EndUserP3, and EndUserP5); and (2) the ability to add the expected language style (EndUserP2, EndUserP3, and EndUserP5).

\subsection{Design Process Validation}

To validate the proxy-user engaged triple diamond design process for AAC systems design, we conjecture that the system outputs consistently meet the expectation of both proxy-users and end-users, as discussed in Section~\ref{section:procedure}. 
Therefore, we compared the keystroke savings and semantic similarity between proxy-users and end-users under the same system settings.

\begin{figure}[H]
    \centering
    \includegraphics[width = 0.6\textwidth]{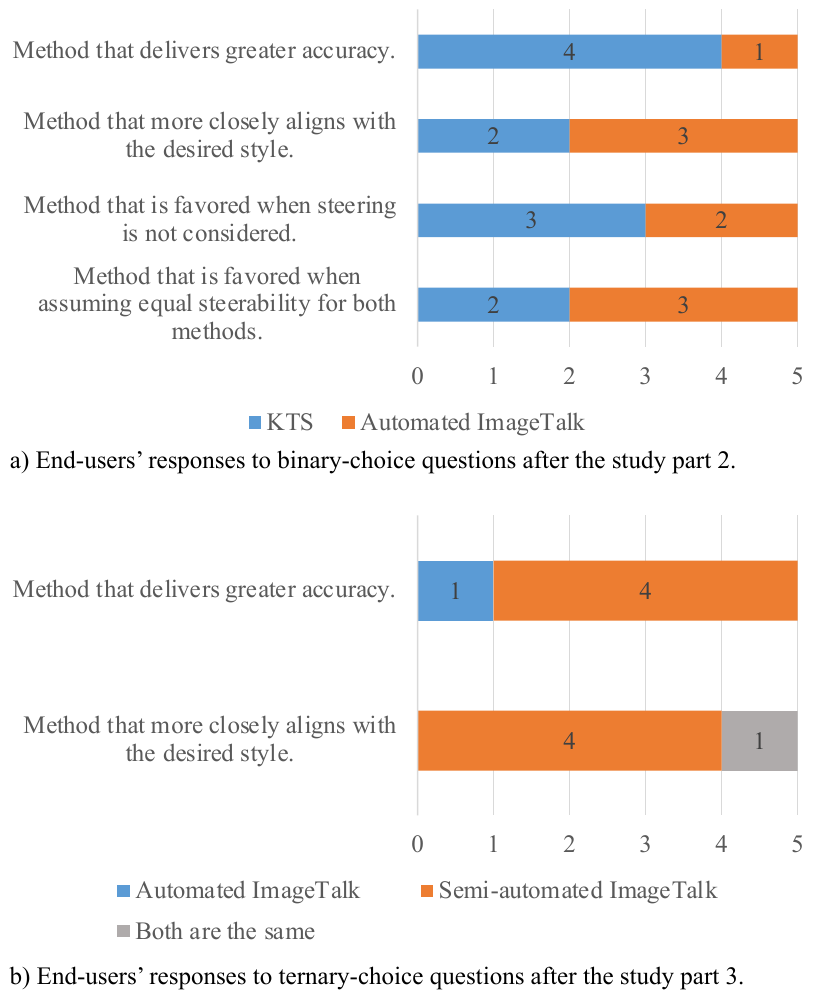}
    \caption{Human evaluation includes two parts in the end-user-centered design phase. The first evaluates end-users' responses to binary-choice questions after study part 2. The second evaluates end-users' responses to ternary-choice questions after study part 3.}
    \label{fig:human_evaluation_stage2}
\end{figure}

Comparing Table~\ref{table:keystroke_savings} with Table~\ref{table:keystroke_savings_stage2}, and comparing Figure~\ref{fig:semantic_similarities} with Figure~\ref{fig:semantic_similarity_stage2}, the Automated ImageTalk system demonstrates comparable high keystroke savings and high semantic similarity for both proxy-users and end-users with consistent reliability. 
Further, a comparison of Figure~\ref{fig:binary_choice} with Figure~\ref{fig:human_evaluation_stage2} a) also reveals a comparable distribution in the preference of the method delivering desired accuracy, language style, and the overall performance when steering is not considered. 

In addition, the guidelines formulated following the human proxy-based design phase were effectively addressed by the steering functions incorporated into the Semi-automated ImageTalk. 
The validity and utility of these guidelines are further supported by strong participant preference for the Semi-automated ImageTalk system in the end-user evaluations as shown in Figure~\ref{fig:human_evaluation_stage2} b).

\subsection{Level of Acceptance of AI-Generated Content}
\begin{figure}[H]
    \centering
    \includegraphics[width = 0.6\textwidth]{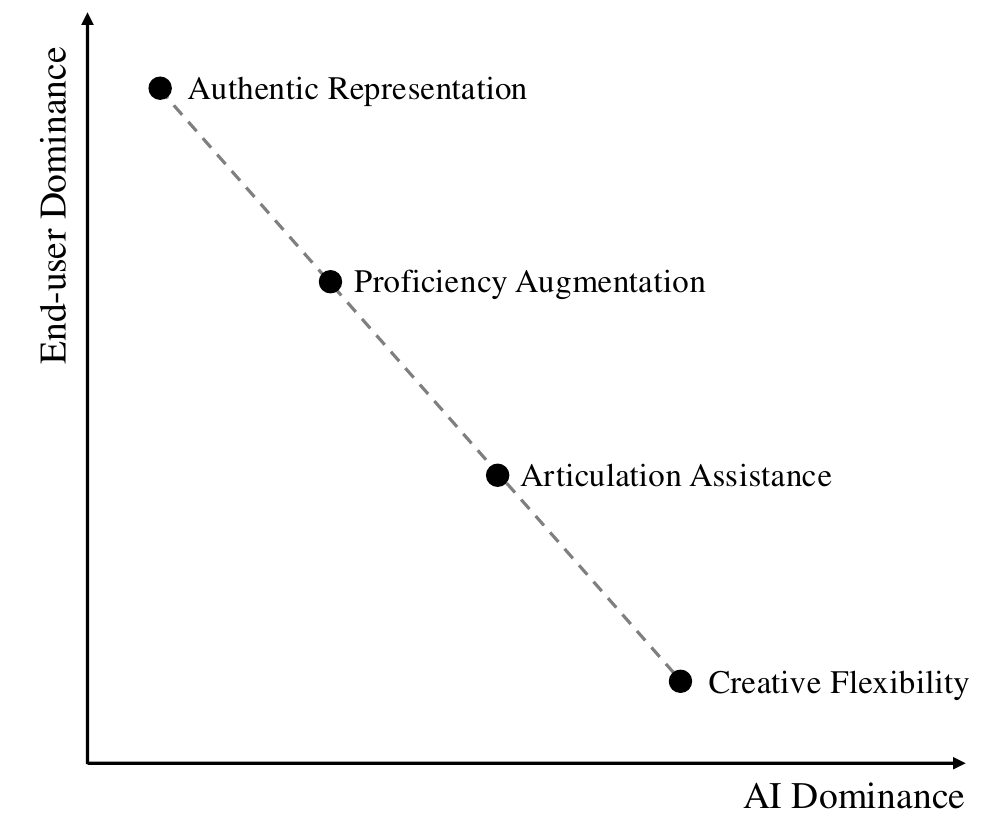}
    \caption{Four levels of acceptance of AI-generated content.}
    \label{fig:level_of_acceptance}
\end{figure}

Diverse language requirements exist among users. We analyzed the end-users' views about the AI-generated narratives generated by both Automated ImageTalk and Semi-automated ImageTalk, and summarized four essential levels of acceptance of AI-generated content. Figure~\ref{fig:level_of_acceptance} illustrates the dynamics of human-AI collaboration and the distribution of end-user acceptance levels regarding AI's role in generating value. In a more end-user-dominant case, AI is expected to reflect the original language of the end-user, while in a more AI-dominant case, creativity is more accepted.
% Add a figure of the spectrum 

\subsubsection{Authentic Representation: Allowing AI to generate content that accurately reflects people's characters and intentions}
This is specified by three aspects: 

for \textbf{content accuracy}, end-users describe the AI-generated stories are: 
\begin{quote}
    \textit{``deviated from the facts''} -- EndUserP1, EndUserP3, EndUserP5 (Automated ImageTalk)
\end{quote}
\begin{quote}
    \textit{``reflect the reality''} -- EndUserP1, EndUserP3, EndUserP5 (Semi-automated ImageTalk)
\end{quote}
for \textbf{language proficiency}, they describe the AI-generated stories are:
\begin{quote}
    \textit{``overly rhetoric''} -- EndUserP5 (Automated ImageTalk)
\end{quote}
\begin{quote}
    \textit{``using complicated words and sentence structures that I won't use''} -- EndUserP2, EndUserP5 (Automated ImageTalk)
\end{quote}
\begin{quote}
    \textit{``using my vocabulary''} -- EndUserP5 (Semi-automated ImageTalk)
\end{quote}
and for \textbf{language style}, they like the AI-generated stories that are:
\begin{quote}
    \textit{``detailed and vivid''} -- EndUserP2, EndUserP3, EndUserP5 (Automated ImageTalk and Semi-automated ImageTalk)
\end{quote}
\begin{quote}
    \textit{``like my language style when I talk to my family and friends''} -- EndUserP1 (Automated ImageTalk)
\end{quote}
\begin{quote}
    \textit{``like my daily language style''} -- EndUserP3 (Semi-automated ImageTalk)
\end{quote}

\subsubsection{Proficiency Augmentation: Allowing AI to augment their proficiency while maintaining fidelity to people's intentions}
Aside from some end-users expressing their needs for the AI-generated story to reflect their intentions, they like the stories that are also:
\begin{quote}
    \textit{``using advanced language''} -- EndUserP1 (Automated ImageTalk and Semi-automated ImageTalk)
\end{quote}

\subsubsection{Articulation Assistance: Allowing AI to articulate intentions that people struggle to express}
Some end-users recognize that:
\begin{quote}
    \textit{``everyone has language barriers, especially when they are in unfamiliar situations or contexts''} -- EndUserP2
\end{quote}
and they find the AI-generated stories: 
\begin{quote}
    \textit{``articulate the intentions that I struggled to express''} -- EndUserP2 (Automated ImageTalk)
\end{quote}

\subsubsection{Creative Flexibility: Allowing AI to exercise creativity and being receptive to fictional content}
Interestingly, some end-users changed their intentions after they read the initial AI-generated story, even though they did not experience the described event. This is because: 
\begin{quote}
    \textit{``I find it (the initial generated story) very interesting and want to imagine what if I have this experience''} -- EndUserP4 (Automated ImageTalk)
\end{quote}
Accordingly, instead of steering the story to their original intentions, they decide to: 
\begin{quote}
    \textit{``refine it based on AI's creativity''} -- EndUserP4 (Semi-automated ImageTalk)
\end{quote}

\subsection{Steering the Generated Story}
We studied two steering methods during the development phase: (1) utilizing the initially generated story as a new input and employing prompts to guide its direction; and (2) directly manipulating user inputs and internal decisive errors to regenerate narratives.
Our findings indicate that the first approach often struggles to discern the prompt's intentions and refine details adequately, whereas the second approach offers better control over the outcome.

\section{Discussion}
\label{section:discussion}

\subsection{Enhancing Communication Through Storytelling}
Storytelling is an essential part of people's daily conversations.
In the context of this research, stories are not elaborate fictional tales; they are snippets of lived experiences, laced with emotions and unique events. 
These anecdotes form the backbone of many interpersonal interactions. 

However, existing AAC devices either have notably low communication rates~\cite{yusufali2024refining,yusufali2023bridging} or need to follow the predefined narrative structure and/or predefined environment~\cite{reiter2009using,black2010using,black2011amobile,mahmud2010xtag,rodil2018sharing}, which hinder the maintenance of a natural communication flow for AAC users~\cite{light1988interaction}. 
This can lead to learned helplessness and develop a passive communicative style in the long term~\cite{basil1992social}. 

Therefore, when devising an optimal AAC system, it becomes imperative to ensure that users can weave these stories quickly, yet with enough depth to maintain the conversation's richness.

The ImageTalk system addresses this by harnessing the inherent richness of images. As visual aids, images encapsulate complex emotions and events with an immediacy that text often struggles to achieve. Combine this with the prowess of LLMs, which are adept at generating coherent text from limited information.
Specifically, this system allows users to input photographs and keywords, subsequently generating a story that reflects the context information extracted from the provided photographs and keywords. 

In general, Semi-automated ImageTalk demonstrates substantial keystroke savings of 95.6\%, marked by consistent performance and high user satisfaction, signifying a significant 21\% improvement in keystroke savings and 4.5 times the consistency when compared to the KTS approach, as measured by the standard deviation of keystroke savings, which solely relies on keywords for narrative generation. 
These substantial keystroke savings hold immense potential for enhancing the text entry rates of AAC users. 
However, the pivotal question remains: how can this potential be effectively translated into tangible efficiency for end-users? This question can be addressed in future work by further completing the user experience and user interface design. 

\subsection{Design Guidelines}
In the \emph{proxy-user-based design phase}, quantitative analyses consistently reveal that Automated ImageTalk maintains higher semantic similarity compared to KTS. 
However, qualitative analyses present intriguing disparities. 
KTS is perceived to exhibit superior accuracy, while Automated ImageTalk emerges as the more favored choice in scenarios where the generated narrative can be further edited, as evaluated by ten proxy-users without speech and mobility difficulties.

This interesting revelation gives rise to three essential design guidelines for the human-in-the-loop development of AI-powered text generation systems:
\begin{enumerate}
    \item Support fine-grained language style control.
    \item Detect potential decisive errors in the internal information flow of the system.
    \item Support fast amendment of decisive errors in system output.
\end{enumerate}

Consequently, to address these design guidelines, we introduced a steering function in the \emph{end-user-centered design phase}, which substantially improved user satisfaction with AI-generated narratives. 
In addition, we distilled four levels of acceptance to AI-generated content for AAC users, from low to high:
\begin{description}
    \item[Authentic Representation] Allowing AI to generate content that accurately reflects people's characters and intentions. 
    \item[Proficiency Augmentation] Allowing AI to augment people's proficiency while maintaining fidelity to people's intentions. 
    \item[Articulation Assistance] Allowing AI to articulate intentions that people struggle to express. 
    \item[Creative Flexibility] Allowing AI to exercise creativity and being receptive to fictional content. 
\end{description}

\section{Conclusion}
\label{section:conclusion}
In summary, we iteratively designed, developed, and evaluated a multimodal AAC text generation system called ImageTalk, which is driven by image recognition and natural language generation in tandem. Our results reveal a highly promising performance potential in terms of keystroke savings but also reveal some challenges in allowing end-users to fully benefit from such performance improvements. 
Our analysis of the quantitative and qualitative data of the study allowed us to distill three guidelines and four levels of acceptance of AI-generated content to guide future research. 
To better meet the operational needs of the end-users, our future work will focus on the interaction design of ImageTalk. This includes refining interaction methods of image selection, keyword input, text editing, and language style selection. Considering the motor difficulties of end-users, the high keystroke savings provide an advantageous space for these interaction designs.
We hope these findings will stimulate further research and engage more researchers in this promising area of AAC. To help other AAC researchers and designers, we will release the source code of the system.

\section*{Open Science}
To further aid AAC design and research, We provide the implementation of ImageTalk as open-source to facilitate further exploration of multimodal interaction systems. The code can be accessed at:
\url{https://github.com/boyiny/ImageTalk}. 

\section*{Acknowledgment}
Special thanks to Dr. Su-Jing Wang and Qin Sun from the Institute of Psychology, Chinese Academy of Sciences, for helping coordinate the participants (plwMND) and arranging user studies. 

\bibliographystyle{ACM-Reference-Format}
\bibliography{sample-base}
\end{document}